\documentclass [a4paper,12pt] {article}

\usepackage{latexsym}
\usepackage{amsfonts}
\usepackage{amsmath}
\usepackage{amssymb}

  \def\pp{{\mathchoice
          {
              \kern 1pt%
              \raise 1pt
              \vbox{\hrule width5pt height0.4pt depth0pt
                    \kern -2pt
                    \hbox{\kern 2.3pt
                          \vrule width0.4pt height6pt depth0pt
                          }
                    \kern -2pt
                    \hrule width5pt height0.4pt depth0pt}%
                    \kern 1pt
           }
            {
              \kern 1pt%
              \raise 1pt
              \vbox{\hrule width4.3pt height0.4pt depth0pt
                    \kern -1.8pt
                    \hbox{\kern 1.95pt
                          \vrule width0.4pt height5.4pt depth0pt
                          }
                    \kern -1.8pt
                    \hrule width4.3pt height0.4pt depth0pt}%
                    \kern 1pt
            }
            {
              \kern 0.5pt%
              \raise 1pt
              \vbox{\hrule width4.0pt height0.3pt depth0pt
                    \kern -1.9pt  
                    \hbox{\kern 1.85pt
                          \vrule width0.3pt height5.7pt depth0pt
                          }
                    \kern -1.9pt
                    \hrule width4.0pt height0.3pt depth0pt}%
                    \kern 0.5pt
            }
            {
              \kern 0.5pt%
              \raise 1pt
              \vbox{\hrule width3.6pt height0.3pt depth0pt
                    \kern -1.5pt
                    \hbox{\kern 1.65pt
                          \vrule width0.3pt height4.5pt depth0pt
                          }
                    \kern -1.5pt
                    \hrule width3.6pt height0.3pt depth0pt}%
                    \kern 0.5pt
            }
        }}

  \def\mm{{\mathchoice
                       {
                             \kern 1pt
               \raise 1pt    \vbox{\hrule width5pt height0.4pt depth0pt
                                  \kern 2pt
                                  \hrule width5pt height0.4pt depth0pt}
                             \kern 1pt}
                       {
                            \kern 1pt
               \raise 1pt \vbox{\hrule width4.3pt height0.4pt depth0pt
                                  \kern 1.8pt
                                  \hrule width4.3pt height0.4pt depth0pt}
                             \kern 1pt}
                       {
                            \kern 0.5pt
               \raise 1pt
                            \vbox{\hrule width4.0pt height0.3pt depth0pt
                                  \kern 1.9pt
                                  \hrule width4.0pt height0.3pt depth0pt}
                            \kern 1pt}
                       {
                           \kern 0.5pt
             \raise 1pt  \vbox{\hrule width3.6pt height0.3pt depth0pt
                                  \kern 1.5pt
                                  \hrule width3.6pt height0.3pt depth0pt}
                           \kern 0.5pt}
                       }}

\catcode`@=11
\def\un#1{\relax\ifmmode\@@underline#1\else
        $\@@underline{\hbox{#1}}$\relax\fi}
\catcode`@=12

\let\du=\du                     
\let\br=\ub                     

\def\a{\alpha}
\def\b{\beta}
\def\c{\chi}
\def\d{\delta}

\def\f{\phi}
\def\g{\gamma}

\def\l{\lambda}
\def\m{\mu}
\def\n{\nu}

\def\p{\pi}
\def\q{\theta}

\def\s{\sigma}
\def\t{\tau}

\def\F{\Phi}

\def\L{\Lambda}

\def\Q{\Theta}

\def\ve{\varepsilon}

\def\ct{{\cal T}}

\def\bo{{\raise-.5ex\hbox{\large$\Box$}}}               
\def\pa{\partial}                                       
\def\de{\nabla}                                         
\def\TH{{\raise.2ex\hbox{$\displaystyle \bigodot$}\mskip-4.7mu \llap H \;}}
\def\face{{\raise.2ex\hbox{$\displaystyle \bigodot$}\mskip-2.2mu \llap {$\ddot
        \smile$}}}                                      

\def\Bar#1{\overline{#1}}                       
\def\ket#1{\left| #1\right\rangle}              
\def\VEV#1{\left\langle #1\right\rangle}        
\def\leftrightarrowfill{$\mathsurround=0pt \mathord\leftarrow \mkern-6mu
        \cleaders\hbox{$\mkern-2mu \mathord- \mkern-2mu$}\hfill
        \mkern-6mu \mathord\rightarrow$}
\def\dvec#1{\vbox{\ialign{##\crcr
        \leftrightarrowfill\crcr\noalign{\kern-1pt\nointerlineskip}
        $\hfil\displaystyle{#1}\hfil$\crcr}}}           
\def\dt#1{{\buildrel {\hbox{\LARGE .}} \over {#1}}}     

\def\frac#1#2{{\textstyle{#1\over\vphantom2\smash{\raise.20ex
        \hbox{$\scriptstyle{#2}$}}}}}                   
\def\half{\frac12}                                        
\def\sfrac#1#2{{\vphantom1\smash{\lower.5ex\hbox{\small$#1$}}\over
        \vphantom1\smash{\raise.4ex\hbox{\small$#2$}}}} 
\def\bfrac#1#2{{\vphantom1\smash{\lower.5ex\hbox{$#1$}}\over
        \vphantom1\smash{\raise.3ex\hbox{$#2$}}}}       
\def\afrac#1#2{{\vphantom1\smash{\lower.5ex\hbox{$#1$}}\over#2}}    
\def\on#1#2{\mathop{\null#2}\limits^{#1}}               
\def\bvec#1{\on\leftarrow{#1}}                  

\def\[{\lfloor{\hskip 0.35pt}\!\!\!\lceil}
\def\]{\rfloor{\hskip 0.35pt}\!\!\!\rceil}

\def\du#1#2{_{#1}{}^{#2}}
\def\ud#1#2{^{#1}{}_{#2}}

\def\fracm#1#2{\hbox{\large{${\frac{{#1}}{{#2}}}$}}}
\def\ha{{\fracmm12}}

\def\Tr{{\rm Tr}}

\def\un{\underline}
\def\fracmm#1#2{{{#1}\over{#2}}}

\def\low#1{{\raise -3pt\hbox{${\hskip 0.75pt}\!_{#1}$}}}

\def\Dot#1{\buildrel{_{_{\hskip 0.01in}\bullet}}\over{#1}}
\def\dt#1{\Dot{#1}}

\newskip\humongous \humongous=0pt plus 1000pt minus 1000pt
\def\caja{\mathsurround=0pt}
\def\eqalign#1{\,\vcenter{\openup2\jot \caja
        \ialign{\strut \hfil$\displaystyle{##}$&$
        \displaystyle{{}##}$\hfil\crcr#1\crcr}}\,}
\newif\ifdtup

\def\np#1#2#3{Nucl.~Phys.~{\bf B{#1}} (19{#2}) #3}

\parskip=\medskipamount                 
\thicklines                         

\begin{document}
\thispagestyle{empty}

{\hbox to\hsize{
\vbox{\noindent February 2005 \hfill hep-th/0502026 }}}

\noindent
\vskip1.3cm
\begin{center}

{\Large\bf Non-anticommutative Deformation of Effective
\vglue.1in
           Potentials in Supersymmetric Gauge Theories~\footnote{
Supported in part by the Japanese Society for Promotion of Science (JSPS)}}
\vglue.2in

Tomoya Hatanaka~\footnote{Email address: thata@kiso.phys.metro-u.ac.jp},
Sergei V. Ketov~\footnote{Email address: ketov@phys.metro-u.ac.jp},
Yoshishige Kobayashi~\footnote{Email address: yosh@phys.metro-u.ac.jp},
and Shin Sasaki~\footnote{Email address: shin-s@phys.metro-u.ac.jp}

{\it Department of Physics, Faculty of Science\\
     Tokyo Metropolitan University\\
     1--1 Minami-osawa, Hachioji-shi\\
     Tokyo 192--0397, Japan}
\end{center}
\vglue.2in
\begin{center}
{\Large\bf Abstract}
\end{center}

\noindent
We studied a nilpotent Non-Anti-Commutative (NAC) deformation of the 
effective superpotentials in supersymmetric gauge theories, caused by
a constant self-dual graviphoton background. We derived the simple 
non-perturbative formula applicable to any NAC (star) deformed chiral 
superpotential. It is remarkable that the deformed superpotential is always 
`Lorentz'-invariant. As an application, we considered the NAC deformation of
the pure super-Yang-Mills theory whose IR physics is known to be described by 
the Veneziano-Yankielowicz superpotential  (in the undeformed case). The 
unbroken gauge invariance of the deformed effective action gives rise to 
severe restrictions on its form. We found a non-vanishing gluino condensate in
vacuum but no further dynamical supersymmetry breaking in the deformed theory.
   
\newpage

\section{Introduction}

It is widely believed that four-dimensional N=1 supersymmetric gauge theories
are relevant to the real world physics, {\it cf.} the Minimal Supersymmetric
Standard Model (MSSM). Amongst the important unsolved issues are confinement,
a mass gap, chiral- and super-symmetry breaking.  They are all related to 
the well known fact that the effective description of infra-red physics in a 
non-abelian gauge theory is usually a strong coupling problem. In the absence
of an analytic proof of confinement and chiral symmetry breaking, it is quite
natural to assume them and then figure out the effective action, as is usual in
 the standard QCD. For a review of the nonperturbative dynamics in the
supersymmetric gauge theories, see ref.~\cite{is}. Our motivation behind this 
paper was a search for a dynamical supersymmetry breaking by certain 
supergravitational corrections to the supersymmetric gauge theory, 
caused by non-anticommutativity (NAC) of the fermionic coordinates in 
superspace \cite{bs,kpt,ovr}. To the best of our knowledge, even an exact 
(non-perturbative) NAC deformation of the non-polynomial effective potentials 
was never calculated in the past, which is the necessary pre-requisite to 
their truly non-perturbative physical applications. The 
NAC itself is known to break supersymmetry by half \cite{sei}, so our problem
here is first to compute the NAC-deformation of any superpotential and then to
study whether supersymmetry could be dynamically broken.

Details are dependent upon the matter content of a gauge theory. In this paper
we only consider the {\it pure} N=1 Super-Yang-Mills (SYM) theory with an 
$U(N_c)=SU(N_c)\times U(1)$ gauge group, i.e. without matter~.\footnote{We 
postpone adding a supersymmetric matter to another publication \cite{2pap}.} 
In the IR limit the $SU(N_c)$ confines while the $U(1)$ is weakly coupled. The 
field content of the simplest confining theory (called the N=1 supersymmetric 
gluodynamics) is given by gluons and their fermionic partners -- gluinos,
all in the adjoint representation of the gauge group. The low-energy effective
action of that theory is known since 1982 \cite{vy}. 

Our paper is organized as follows. In sect.~2 we give our notation, as 
regards the quantum N=1 SYM theory. Sect.~3 is devoted to a brief 
introduction into the Non-Anti-Commutativity (NAC) along the lines of 
ref.~\cite{sei}. It shows our setup, where we also add our comments about the
NonAntiCommutativity (NAC) versus the usual NonCommutativity in field theory. 
The main part of our work is given by Sect.~4 where we explicitly compute the 
non-perturbative NAC-deformation of an arbitrary chiral superpotential, and
apply our results to the standard Veneziano-Yankielowicz (VY) effective
superpotential \cite{vy}. Our conclusions are given by Sect.~5. Some notation
and technical details of our calculation are collected in two Appendices, A and
B.

\newpage

\section{N=1 SYM in components and in superspace}

The standard fundamental action of the N=1 SYM theory in components reads
$$ I_{\rm SYM} = -\fracmm{1}{2g^2} \int \Tr \left[ F\wedge {}^*F 
- 4i\bar{\l}\tilde{\s}^{\m}\de_{\m}\l\right]
 +\fracmm{\Q}{16\p^2}\int \Tr\, F\wedge F~,
\eqno(2.1)$$ 
where we have introduced the YM Lie algebra-valued two-form $F=F_{\m\n}dx^{\m}
\wedge dx^{\n}$ and the chiral spinors, $\l$ and $\bar{\l}$. The ${}^*F$
stands for the Poincar\'e-dual two-form of $F$. We use the standard (in
supersymmetry) two-component notation for spinors \cite{wb}. It is common to
unify the YM coupling constant $g$ and the theta-parameter $\Q$ into a single
(complex) coupling constant
$$ \t = {\rm Re}\,\t + i\,{\rm Im}\,\t=\fracmm{4\p}{g^2} -i\fracmm{\Q}{2\p}~~.
\eqno(2.2)$$ 
We study either {\it Euclidean} or {\it Atiyah-Ward} versions of the SYM 
theory in $4+0$ or $2+2$ dimensions, respectively~\cite{kgn}. The 
last term in eq.~(2.1) can be identified with the quantized instanton charge
(in Euclidean space) 
$$ Q_{\rm instanton}=\fracmm{1}{16\p^2}\int \Tr\, F\wedge F \in {\bf Z}~~.
\eqno(2.3)$$

The `Lorentz' group factorizes in Euclidean space, $SO(4)\cong SU(2)_L\times 
SU(2)_R$, as well as in Atiyah-Ward space, $SO(2,2)\cong SU(1,1)_L\times 
SU(1,1)_R$, so that the `left' and `right' chiral spinors, $\l$ and
$\bar{\l}$, are fully independent upon each other (being also real in  
$2+2$ dimensions) \cite{kgn}. 

Since the theory is supersymmetric, it is better to make supersymmetry manifest
and thus make sure that quantum theory is also supersymmetric. It can be done
in superspace $(x^{\m},\q^{\a},\bar{\q}^{\dt{\a}})$, where $\m=1,2,3,4$ and
$\a=1,2$. Here $\q^{\a}$ and $\bar{\q}^{\dt{\a}}$ are the anticommuting
(Grassmann) spinor coordinates of superspace, whereas  $x^{\m}$ stand for the 
usual bosonic (commuting) coordinates in ${\bf R}^4$.

The superspace extension of a YM field $A_{\m}$ is given by a general, Lie 
algebra-valued N=1 scalar superfield $V(x,\q,\bar{\q})$ subject to the 
supergauge transformations,
$$ e^V \;\to\; e^{V'}=e^{-i\bar{\L}}e^V e^{i\L}~~,\eqno(2.4)$$
with the gauge parameter $\L$ being a chiral superfield, i.e.  
$\bar{D}_{\dt{\a}}\L=0$. The spinorial supercovariant derivatives in 
superspace, $D\low{\a}$ and $\bar{D}_{\dt{\a}}$, are supposed to commute with
spacetime translations and supersymmetry generators, by their definition.  
See ref.~\cite{wb} and Appendix A for more details. 

The superfield analogue of the YM field strength is given by the N=1 chiral
gauge-covariant superfield strength (of canonical dimension $3/2$),
$$ W_{\a} = -\fracmm{1}{4}\Bar{D}^2\left(e^{-V}D_{\a}e^V\right)~,\eqno(2.5a)$$
and its N=1 anti-chiral cousine,
$$\bar{W}_{\dt{\a}} = 
\fracmm{1}{4}D^2\left(e^{V}\bar{D}_{\dt{\a}}e^{-V}\right)~.\eqno(2.5b)$$
It is useful  to introduce a chiral basis in superspace, where 
chirality is manifest, by shifting the bosonic coordinates as
$$ y^{\m}=x^{\m} + i\q^{\a}\s^{\m}_{\a\dt{\a}}\bar{\q}^{\dt{\a}}~~.\eqno(2.6)$$
Then any chiral superfield $\F$ is simply a function $\F(y,\q)$. The spinorial
covariant derivatives in the chiral basis are given by
$$ D_{\a}=\fracmm{\pa}{\pa\q^{\a}}+2i\s^{\m}_{\a\dt{\a}}\bar{\q}^{\dt{\a}}
\fracmm{\pa}{\pa y^{\m}}~~,\quad \bar{D}_{\dt{\a}}=-
\fracmm{\pa}{\pa\bar{\q}^{\dt{\a}}}~~~.\eqno(2.7)$$ 
They obey an algebra
$$ \{D_{\a},D_{\b}\}=0~,\quad
 \{\bar{D}_{\dt{\a}},\bar{D}_{\dt{\b}}\}=0~,\quad
\left\{ D\low{\a},\bar{D}_{\dt{\a}}\right\}=
-2i\s^{\m}_{\a\dt{\a}}\pa_{\m}~~.\eqno(2.8)$$

The N=1 SYM action in superspace reads
$$\eqalign{
 I_{SYM,~s}  &~= \fracmm{\t}{16\p}\int d^4y d^2\q\,\Tr\, W^{\a}W_{\a}+
\fracmm{\bar{\t}}{16\p}\int d^4\bar{y} d^2\bar{\q}\,\Tr\,\bar{W}_{\dt{\a}}
\bar{W}^{\dt{\a}} \cr
&~= I_{SYM} + \fracmm{1}{g^2}\int d^4x \Tr\, D^2~, \cr}\eqno(2.9)$$
where $D$ is the auxiliary field (of dimension $2$) needed to close the algebra
of the  supersymmetry transformations on the SYM fields, while $I_{SYM}$ is 
given by eq.~(2.1). The auxuliary field $D$ decouples, while its equation of
motion is $D=0$ (this feature is not so obvious when considering the highly
non-linear effective actions in superspace).

The gauge freedom (2.4) can be used to get rid of some of the field components
of the gauge superfield $V$, by putting it into the form
$$ V(y,\q,\bar{\q})=-(\q\s^{\m}\bar{\q})A_{\m}(y) +i\q^2\bar{\q}_{\dt{\a}}
\bar{\l}^{\dt{\a}}(y) -i\bar{\q}^2\q^{\a}\l_{\a}(y)+\fracm{1}{2}\q^2\bar{\q}^2
[D(y)-i\pa\cdot A(y)]~,\eqno(2.10)$$
without breaking supersymmetry, thus rendering $V$ to obey the nilpotency 
condition $V^3=0$. This is known as the Wess-Zumino gauge \cite{wb}. 
Substituting (2.10) into (2.5a) yields
$$ W_{\a}(y,\q)=-i\l_{\a}+\left[\d^{\b}_{\a}D-i(\s^{\m\n})_{\a}{}^{\b}F_{\m\n}
\right]\q_{\b} +\q^2\s^{\m}_{\a\dt{\a}}\de_{\m}\bar{\l}^{\dt{\a}}~~.
\eqno(2.11)$$

The classical SYM action is not only supersymmetric, it is also scale and
chirally invariant, because of the absence of dimensional parameters and the
`left'-`right' symmetry. The left-right symmetry commutes with supersymmetry, 
while it should not be confused with the chiral R-symmetry \cite{wb} that does
 not commute with supersymmetry. In quantum theory, the scale invariance and 
the R-symmetry are broken due to anomalies. Supersymmetry is expected to be 
preserved, while the left-right symmetry is expected to be violated too.

The one-loop renormalization group (RG) beta-function of the theory (2.1) can 
be computed by the standard procedure of quantum field theory, with the well 
known result
$$ \b(g)= \m\fracmm{dg}{d\m}=-\fracmm{3N_cg^3}{16\p^2}~~,\eqno(2.12)$$
where the RG scale $\m$ and the running coupling constant $g(\m)$ have been 
introduced. The negative sign on the r.h.s. of eg.~(2.12) implies the UV
asymptotic freedom as well as the strong coupling in the IR limit. The 
anomalous trace of the stress-energy tensor and the anomalous divergence of
the axial current are also well known (see e.g., ref.~\cite{vy}),
$$ T^{\m}_{\m}=\fracmm{\b(g)}{2g}(F_{\m\n}^a)^2~,\eqno(2.13a)$$
and
$$ \pa_{\m}J_5^{\m}=-\fracmm{\b(g)}{2g}F_{\m\n}^a{}^*F_{\m\n}^a~~.
\eqno(2.13b)$$
The supercurrent $(S_{\m}^{\a},\bar{S}^{\dt{\a}}_{\m})$ is conserved, but it
is subject to the superconformal anomaly,
$$\tilde{\s}^{\m\dt{\a}\b}S_{\m\b}=\fracmm{\b(g)}{g}F_{\m\n}^a
(\tilde{\s}_{\m\n})\ud{\dt{\a}}{\dt{\g}}\bar{\l}^{a\dt{\g}}~,\eqno(2.13c)$$
and similarly for $\bar{S}^{\dt{\a}}_{\m}$. Both the currents and their 
anomalies are known to form supermultiplets \cite{fz,sspace}. In particular, 
the stress-tensor $T\du{\m}{\n}$, the axial current $J^{\m}_5$  and the
supercurrent $S^{\a}_{\m}$ belong to the field components of a constrained 
vector superfield $T_{\a\dt{\a}}$ subject to the classical relations  
$$ D^{\a}T_{\a\dt{\a}}=\bar{D}^{\dt{\a}}T_{\a\dt{\a}}=0~.\eqno(2.14)$$
The anomalies form a chiral superfield $\ct$ with $\bar{D}_{\dt{\a}}\ct=0$ 
\cite{sspace}. The classical relations (2.14) get modified in the quantum SYM 
theory as
$$\bar{D}^{\dt{\a}}T_{\a\dt{\a}} \propto D_{\a}\ct~,\quad {\rm and}\quad  
D^{\a}T_{\a\dt{\a}} \propto \bar{D}_{\dt{\a}}\bar{\ct}~~.\eqno(2.15)$$
In part, the anticommutator of a supersymmetry charge $Q$ with the 
supercurrent results in the conformal anomaly proportional to $F^2$. Taking 
the vacuum expectation value of that relation implies that a non-vanishing 
value of $\VEV{F^2}$ gives rise to $\VEV{T^{\m}_{\m}}\neq 0$ and 
$Q\ket{0}\neq 0$, i.e. it results in a {\it Dynamical Supersymmetry Breaking} 
(DSB).  See a review \cite{ff} for more details about the DSB.

In supersymmetry $\Tr\,F^2$ and $\Tr\,{}^*FF$ are united into a 
complex field that belongs to a chiral supermultiplet, together with the 
gaugino composite field $\Tr\,(\l^{\a}\l_{\a})$. It is known as the N=1 chiral
{\it glueball} superfield (all traces here are taken in the $SU(N_c)$)
$$ S \propto \Tr\,(W^{\a}W_{\a})~.\eqno(2.16)$$
The r.h.s. of eq.~(2.16) is of mass dimension $3$, while eq.~(2.13) gives the
natural normalization factor $\b(g)/2g$, as in ref.~\cite{vy}. 
We are going to use a {\it dimensionless} glueball superfield $S$ by
further rescaling $S\to S/\m^3$, where $\m$ is the RG scale.   

Next, though the classical glueball superfield is nilpotent due to the 
fermionic statistics of gluions, $S^{N_c^2}=0$, it is not
necessarily true in quantum theory since the nilpotency condition is subject
to quantum corrections. We ignore this subtlety in what follows.

The field components of the glueball superfield $S$ are given by 
(up to a constant)
$$ \Tr(\l^2)\equiv \f~,\quad 
 \Tr\left[ \fracm{i}{2}(\s^{\m\n})\du{\a}{\b}F_{\m\n}\l_{\b}
\right]\equiv\c_{\a}~,\quad 
 \Tr(F_{\m\n}F^{\m\n}+iF_{\m\n}{}^*F^{\m\n})\equiv M. \eqno(2.17)$$    

The glueball superfield $S$ is a singlet (colorless) with respect to the gauge 
group, so that it appears to the natural {\it order} parameter in any 
description of the IR physics of the N=1 supersymmetric gluodynamics. It is 
usually assumed that the IR physics can be described by some effective action 
depending upon the chiral glueball superfield $S$ and its anti-chiral cousine 
$\bar{S}$, while both are to be considered as the independent superfields in 
the IR limit \cite{vy,cdsw}. In particular, $M$ is going to play the role of 
the auxiliary field in what follows, while the DSB occurs whenever 
$\VEV{M}\neq 0$.

The most general, manifestly supersymmetric low-energy effective action 
(without higher derivatives of the field components in eq.~(2.17)) is given by
$$ I[S,\bar{S}] = \m^2
\int d^4x d^2\q d^2\bar{\q}\, K(S,\bar{S})+ \m^3\int d^4 yd^2\q\, V(S) +
\m^3 \int d^4\bar{y} d^2\bar{\q}\,\bar{V}(\bar{S})~,\eqno(2.18)$$
where the dimensionless kinetic function $K(S,\bar{S})$ is called a 
{\it K\"ahler} potential, and another dimensionless function $V(S)$ is called 
a {\it superpotential}. In Minkowski spacetime the function $\bar{V}(\bar{S})$
 is simply the Hermitean conjugate of $V(S)$, though it is not going to be the
 case in sects.~3 and 4, where either Euclidean or Atiyah-Ward spacetime 
signatures are assumed.

As was first discovered in ref.~\cite{vy}, there exists a unique 
non-perturbative scalar superpotential (nowadays famously known as the VY 
superpotential) that reproduces the anomaly structure of the N=1 SYM and that
of the N=1 gluodynamics, namely,
$$ V_{\rm VY}(S)= N_c S\ln S + \t_{\rm ren}S~,\eqno(2.19)$$
where we have introduced the renormalized value $\t_{\rm ren}$ of the SYM 
coupling constant at the scale $\m$. For instance, in one-loop we have
$$ \t_{\rm ren}= \t_0 +3N_c\ln\fracmm{\m}{\m_0}~,\eqno(2.20)$$
with $\m_0$ being the scale where the bare coupling $\t_0$ is defined. It is 
worth mentioning that no dimensional quantities appear in eq.~(2.19). There
are other ways of `deriving' the VY superpotential, either from the field 
theory \cite{cdsw,agy} or from the matrix models \cite{kyoto} by using the 
Dijkgraaf-Vafa correspondence \cite{dv}.

Minimizing the VY superpotential, $V'(S)=0$, one finds a non-vanishing
gluino condensate,
$$\VEV{\Tr\l^2}\propto  \VEV{S}\propto e^{-\t_{\rm ren}/N_c}\propto 
e^{-4\p^2/g^2N_c}~,\eqno(2.21)$$
but no dynamical susy breaking (DSB)  because of $\VEV{M}=\VEV{\Tr F^2}=
\VEV{\Tr \,F{}^*F}=0$.

\section{Nilpotent NAC deformation}

The Non-Anti-Commutative (NAC) deformation of the N=1 superspace is given by
\cite{bs,kpt} 
$$ \{ \q^{\a},\q^{\b} \}=C^{\a\b}~~,\eqno(3.1)$$
where $C^{\a\b}$ are some constants, i.e. the chiral spinorial superspace 
coordinates are no longer Grassmann but satisfy a Clifford algebra (3.1). 
It is consistent to keep unchanged the rest of the (anti)commutation relations
 between the N=1 superspace coordinates (in the chiral basis),
$$ \[ y^{\m},y^{\n}\] = \[ y^{\m},\q^{\a}\] = \[ y^{\m},\bar{\q}^{\dt{\a}}\]=
0~~,\eqno(3.2)$$
as well as
$$  \{ \q^{\a},\bar{\q}^{\dt{\b}} \}= 
\{ \bar{\q}^{\dt{\a}},\bar{\q}^{\dt{\b}} \}=0~~ ,\eqno(3.3)$$
either in Euclidean or Atiyah-Ward spacetime where $\q^{\a}$ and 
$\bar{\q}^{\dt{\a}}$ are truly independent. This choice of a NAC deformation 
is sometimes called {\it nilpotent} \cite{ilz,ks3} since it gives rise to a 
{\it local} deformed field theory. The physical significance of the NAC 
deformation (3.1) in string theory was uncovered by Ooguri and Vafa 
\cite{ovr}. They argued that the $C^{\a\b}$ can be thought of as the vacuum 
expectation values of the self-dual graviphoton field strength 
$F_{\rm graviphoton}^{\m\n}$ (see also refs.~\cite{stony,sei}) with
$$ (\a')^2F^{\a\b}_{\rm graviphoton}=C^{\a\b}~~.\eqno(3.4)$$
In the Calabi-Yau (CY) compactified type-IIB superstrings a self-dual RR-type
5-form can have a non-vanishing flux over certain CY cycles, which gives rise
to a non-vanishing self-dual graviphoton flux in four dimensions. From the
viewpoint of the N=1 SYM theory in four dimensions, the deformation (3.1) can 
be thought of as the result of some gravitational corrections coming after
embedding the gauge theory into extended supergravity or superstrings.

The $C^{\a\b}\neq 0$ in eq.~(3.1) explicitly breaks the
four-dimensional `Lorentz' invariance at the fundamental level. The NAC nature
of $\q$'s can be fully taken into account in field theory  by using the
Moyal-Weyl-type star product (functions of $\q$'s are to be ordered) \cite{sei}
 $$ f(\q)\star g(\q)=f(\q)\,
\exp\left(-\fracmm{C^{\a\b}}{2}\fracmm{\bvec{\pa}}{\pa
\q^{\a}}\fracmm{\vec{\pa}}{\pa\q^{\b}}\right)g(\q) \eqno(3.5)$$
that clearly respects the N=1 superspace chirality, so that the star product of
any two chiral or anti-chiral superfields is again a chiral or anti-chiral
superfield, respectively.

The star product (3.5) is polynomial in the deformation 
parameter~,\footnote{See Appendix A for more about our notation.}
$$ f(\q)\star g(\q)=fg +(-1)^{{\rm deg}f}\fracmm{C^{\a\b}}{2}
\fracmm{\pa f}{\pa\q^{\a}}\fracmm{\pa g}{\pa\q^{\b}}-\det\,C
\fracmm{\pa^2 f}{\pa\q^2}\fracmm{\pa^2 g}{\pa\q^2}~~,\eqno(3.6)$$
where we have used the identities
$$ \det C = \fracm{1}{2}\ve_{\a\g}\ve_{\b\d}C^{\a\b}C^{\g\d}
= \fracm{1}{4}(C_{\m\n})^2~~,\eqno(3.7)$$
and the standard notation between the vector and spinor indices, 
$$ C^{\m\n}= C^{\a\b}\ve_{\b\g}(\s^{\m\n})\du{\a}{\g}~~,\eqno(3.8)$$
with $\m,\n=1,2,3,4$ and $\a,\b,\ldots=1,2$. Any value of the scalar $\det C$ 
is obviously `Lorentz'-invariant, while it can be real in $2+2$ dimensions. 

The nilpotent NAC deformation versus the spacetime Non-Commutativity (NC) 
\cite{dn,sew}, described by the relations $\[y^{\m},y^{\n}\]=iB^{\m\n}\neq 0$,
has several advantages. In particular, the nilpotent NAC does not lead to a 
non-local field theory but to a very limited (finite) number of the new 
vertices. This also implies the absence of the UV/IR mixing problem common to
all NC theories. The nilpotent NAC deformation of an abelian supersymmetric 
gauge theory is almost trivial \cite{ks3}, e.g., there are no $U(1)$ monopoles
and instantons there. Of course, there are also serious problems related to 
NAC. For instance, the nilpotent NAC deformation is only possible in Euclidean
or Atiyah-ward spacetimes, it leads to non-Hermitean actions (see below), 
which may cause problems with unitarity. However, a discussion of unitarity in
 the NAC deformed field theories is beyond the scope of this paper.

The NAC deformation (3.1) of the N=1 SYM theory (2.1) with the $U(N_c)$ gauge
group in Euclidean space was considered by Seiberg \cite{sei}. We refer to his
 paper \cite{sei} for details, but we would like to emphasize here some 
aspects of ref.~\cite{sei} that are going to be relevant for our next sect.~4.
In particular, the deformation (3.1) breaks just half of N=1 or 
$(\half,\half)$ supersymmetry, while another half of supersymmetry 
remains  unbroken \cite{sei}.  

The fate of the gauge invariance in a quantized NAC-deformed SYM theory did 
not receive enough attention in ref.~\cite{sei}, and it remains to be a highly
non-trivial issue. When requiring the component fields to transform in the 
standard (undeformed) way under the gauge transformations, the NAC gauge 
superfield in the WZ gauge has to be modified \cite{sei}~,\footnote{All the 
deformed quantities vs. the undeformed ones are marked by the subscript $C$.}
$$ V_C(y,\q,\bar{\q})=V(y,\q,\bar{\q})-\fracm{i}{4}\bar{\q}^2\q^{\a}\ve_{\a\b}
C^{\b\g}\s^{\m}_{\g\dt{\g}} \{ \bar{\l}^{\dt{\g}},A_{\m} \}~,\eqno(3.9)$$
where $V(y,\q,\bar{\q})$ is given by eq.~(2.10).  By construction \cite{sei},
the {\it residual} gauge transformations (keeping the form of eq.~(3.9) 
intact) of the field components $\left(A_{\m},\l^{\a},\bar{\l}^{\dt{\a}},D
\right)$ are $C$-independent, i.e. of the standard form. It is straightforward
to calculate the deformed gauge superfield strengths (in the WZ gauge). One 
finds \cite{sei}
$$ (W_C)_{\a}(y,\q)=W_{\a}(y,\q) +\ve_{\a\g}C^{\g\b}\q_{\b}\bar{\l}^2 
\eqno(3.10)$$
and
$$\eqalign{
 (\Bar{W}_C)_{\dt{\a}}(\bar{y},\bar{\q}) = ~&
 \Bar{W}_{\dt{\a}}(\bar{y},\bar{\q}) -\bar{\q}^2\left(
\fracm{1}{2}C^{\m\n}\{ F_{\m\n},\bar{\l}_{\dt{\a}} \} +C^{\m\n}\left\{ A_{\n},
\de_{\m} \bar{\l}_{\dt{\a}} -\fracm{i}{4} \[ A_{\m},\bar{\l}_{\dt{\a}} \]
\right\} \right. \cr
~& \left. +\fracm{1}{4}(\det C) \left\{ \bar{\l}^2,\bar{\l}_{\dt{\a}}\right\}
\right)~.\cr}\eqno(3.11)$$
Hence, the NAC chiral superfield strength (3.10) is gauge-covariant,
whereas the NAC anti-chiral superfield strength (3.11) is not as long as 
$C\neq 0$. Moreover, both eqs.~(3.9) and (3.11) contain anti-commutators of 
the Lie algebra-valued gauge fields, whose closure restricts the choice of a
gauge group (e.g., the $U(N_c)$ is ok). 

As regards the fundamental SYM Lagrangian in eq.~(2.9),  it is given by a 
linear combination of two terms,
$$ \int d^2\q\, \Tr\,W^2\quad {\rm and}\quad \int d^2\bar{\q}\,\Tr\,
\Bar{W}^2~~.\eqno(3.12)$$
Their NAC deformations are almost the same, up to a total derivative 
in ${\bf R}^4$ \cite{sei},
$$ \int d^2\q\, \Tr(W^2)_C=\int d^2\q\, \Tr\,W^2
-iC^{\m\n}\Tr(F_{\m\n}\bar{\l}^2) +(\det C)\Tr(\bar{\l}^2)^2 \eqno(3.13)$$
and
$$\int d^2\bar{\q}\,(\Tr\Bar{W}^2)_C= \int d^2\bar{\q}\,\Tr\Bar{W}^2
-iC^{\m\n}\Tr(F_{\m\n}\bar{\l}^2) +(\det C)\Tr(\bar{\l}^2)^2 +{\rm 
total~~derivative}~.\eqno(3.14)$$
Our calculation of the total derivative in eq.~(3.14) reveals that it is not 
gauge-invariant, {\it viz.}
$$ {\rm total~~derivative}=\pa_{\m}\left(C^{\m\n}\Tr\,A_{\n}\bar{\l}^2\right)~.
\eqno(3.15)$$

When considering the effective field theory (2.18) originating from the 
fundamental supersymmetric gauge theory, eqs.~(3.13), (3.14) and (3.15) imply 
that only the {\it holomorphic} superpotential can be affected by the 
NAC-deformation, whereas the anti-holomorphic superpotential cannot (up to a 
linear contribution), as long as the gauge invariance is preserved in
quantum theory (see also refs.~\cite{cpot,imaa}). 

As regards quantum properties of the NAC-deformed SYM theory, it was argued
\cite{ren} that it is still renormalizable to all orders of perturbation theory
despite the presence of apparently non-renormalizable (by power counting) 
$C$-dependent interactions. The gauge invariance of a quantized $N=\ha$ SYM 
theory at one-loop was proved in ref.~\cite{jj}. The gauge-invariant RG 
beta-function and the anomalies of the NAC  deformed SYM theory are aparently 
{\it the same} as that of the undeformed theory (see also ref.~\cite{ags} for 
explicit one-loop calculations).

It is also interesting to see what happens to the $U(1)$ factor of the
gauge group $U(N_c)=SU(N_c)\times U(1)$.~\footnote{We are grateful to G. Silva 
and S. Terashima for discussions about this point.} In the superspace approach
 of ref.~\cite{sei} the $U(N_c)$ gauge transformations are NAC-deformed, while
the $U(1)$ factor is necessary for a closure of the gauge algebra. In the 
WZ gauge \cite{sei} the $U(N_c)$ gauge transformations are undeformed (i.e.
$C$-independent), the $U(1)$ and $SU(N_c)$ gauge transformations can be
separated, but there is a coupling between photinos and gluions due to the 
last term in  eq.~(3.13). We assume, however, that the NAC deformed SYM is 
still well-defined in the IR, where the $SU(N_c)$ confines and the $U(1)$ 
is weakly coupled like that in the undeformed theory.

\newpage

%
%

\def\br#1{\left( #1 \right)}                    
\def\Br#1{\left[ #1 \right]}                    
\def\BR#1{ \left\{ #1 \right\} }                

\def\frap#1#2{\frac{\partial #1}{\partial #2}}	  
\def\frapp#1#2{\frac{\partial^2 #1}{\partial #2}} 


\section{NAC-deformed effective potentials}

We begin with our main result of this paper, which is the beautiful formula
describing the non-perturbative NAC (star) deformation of an {\it arbitrary} 
superpotential $V(f)$,
\begin{eqnarray*}
	\begin{aligned}
		\int \!\! d^2\q \,\, V_{\star}(f)
	&	=
		V'(\f) M
		-
		\frac{1}{2}V''(\f) \c^2
	\\
	&\quad
		+
		\sum^{\infty}_{k=1}
		\fracmm{1}{(2k+1)!}
		\br{ -\det C }^{k}
		M^{2k}
		\br{ 
			V^{(2k+1)}(\f) M 
			- 
			\frac{1}{2}V^{(2k+2)}(\f) \c^2
		}
	\\
	&
		=
		\fracmm{1}{2c}
		\BR{
			V\br{\f + c M}
			-
			V\br{\f - c M}
		}
	\\
	&\quad
		-
		\fracmm{\c^2}{4c M}
		\BR{
			V'\br{\f + c M}
			-
			V'\br{\f - c M} 
		}~, \hfill ~~~~~~~~~~~~~~~~~~~~~~~~~~~~~~~~~~(4.1)
	\end{aligned}  
\end{eqnarray*}
where we have used the notation (A.6) for the field components of a chiral
superfield $f(y,\q)$, and have introduced the effective (`Lorentz'-invariant) 
deformation parameter
$$  \sqrt{-\det C}  \equiv c~~.\eqno(4.2)$$ 
The primes denote the derivatives of the function $V$ with respect to its 
argument. The star subscript means that all the products of $f$'s (in Taylor
expansion of $V(f)$) are to be taken by using the star product (3.5).
 
It is worth mentioning that eq.~(4.1) is `Lorentz'-invariant. In particular, 
as regards the purely bosonic terms, eq.~(4.1) yields the
remarkably simple non-perturbative equation,
$$ \left. \int \!\! d^2\q \,\, V_{\star}(f) \right|_{\rm bosonic} = 
\fracmm{1}{2 c}
		\BR{
			V\br{\f + c M}
			-
			V\br{\f - c M}}\eqno(4.3)$$
which clearly shows that the NAC deformation of any potential $V$ just  
amounts to the $M$-dependent splitting of the leading argument of the 
potential (after integrating over the fermionic coordinates).  

Our equation (4.1) agrees with the earlier calculations \cite{cpot} in the
case of a cubic superpotential (i.e. the NAC deformed supersymmetric  
Wess-Zumino model).~\footnote{When preparing our paper for publication we
learned that some perturbative calculations of the \newline ${~~~~~}$ NAC 
star-deformed action (2.18) also appeared in ref.~\cite{buch}.} 
 
We found eq.~(4.1) as a result of our complicated calculations that we now 
briefly describe. It is clearly enough to prove eq.~(4.1) in the case of a 
power -like superpotential, $V(f)=f^p$, with some positive integer $p$. A
straightforward but tedious application of the rule (3.5) by induction yields
$$ f^p_{\star} = f^p + \sum^{ \[ {\frac{p-2}{2}}\] }_{j=0} A^{(p)}_{j+1}~,
\eqno(4.4)$$
where we have introduced the notation
\begin{eqnarray*}
	\begin{aligned}
		A^{(p)}_{j+1} = &
		\br{-\det C}^{j+1}
		\sum^{p-j}_{k_1=1}
		\sum^{p-j-k_1}_{k_2=1}
		\cdots
		\sum^{p-j-\sum^{j}_{r=1}k_r}_{k_{j+1}=1} 
 \fracmm{\pa^2}{\pa\q^2}\br{f^{p-j-\sum^{j+1}_{s=1}k_s}}\times
	\\
	& \times \br{\fracmm{\pa^2 f}{\pa\q^2}}^{j+1}
		\fracmm{\pa^2 f^{k_{j+1}}}{\pa\q^2}
		\fracmm{\pa^2 f^{k_{j}}}{\pa\q^2}
		\cdots
		\fracmm{\pa^2 f^{k_2}}{\pa \q^2}
		f^{k_1-1}~~~.\hfill ~~~~~~~~~~~~~~~~~~~~~~~~~~~~~~~~~~~~(4.5)
	\end{aligned}
\end{eqnarray*}
Integrating over $\q$'s or, equivalently, taking the last field component of 
the chiral superfield (4.5) gives rise to many cancellations, with the result
(e.g., when $p\neq 2j+2$)
\begin{eqnarray*}
	\begin{aligned}
		\int \!\! d^2\q \,\, A^{(p)}_{j+1}	
	&	=
		\br{-\det C}^{j+1}
		\sum^{p-j}_{k_1=1}
		\sum^{p-j-k_1}_{k_2=1}
		\cdots
		\sum^{p-j-\sum^{j}_{r=1}k_r}_{k_{j+1}=1}
	\\
	&
		\int \!\! d^2\q \,\,
		\fracmm{ 
               \br{p-j-\sum^{j+1}_{s=1}k_s} (k_1-1) k_2 \cdots k_{j+1} }
		{ p-2(j+1) }
		f^{p-2(j+1)}
		\br{ \fracmm{\pa^2 f}{\pa \q^2} }^{2(j+1)}~.
 \hfill ~(4.6a)
	\end{aligned} 
\end{eqnarray*}
When $p=2j+2$ and thus even, we found the very simple formula,
$$ A^{(p)}_{p/2}=\left( -\det C\right)^{p/2}\left( \fracmm{\pa^2 f}{\pa\q^2}
\right)^p~,\quad {\rm so~~that} \quad \int d^2\q\, A^{(p)}_{p/2} =0~. 
\eqno(4.6b)$$ 	

After combining eqs.~(4.4), (4.5) and (4.6) we find 
$$  \eqalign{
 \int \!\! d^2\q \, f^p_{\star} = &  \int \!\! d^2\q \, f^p \cr
     & +
		\sum^{\[ {\frac{p-2}{2}} \] }_{j=0}
		\br{-\det C}^{j+1}
		\fracmm{p(p-1)\cdots(p-2j-1)}{(2j+3)!}
		\int \!\! d^2\q \, 
		f^{p-2(j+1)}
		\br{ \fracmm{\pa^2 f}{\pa\q^2} }^{2(j+1)},\cr} \eqno(4.7)$$
where we have also used the crucial combinatorial identity
$$\sum_{k_1=1}^{p-j}  
	\sum_{k_2=1}^{p-j-k_1} 
	\cdots
	\sum_{k_{j+1}=1}^{p-j-\sum_{r=1}^j k_r}
	(p-j-\sum_{s=1}^{j+1} k_s)(k_1-1)k_2 \cdots k_{j+1}
	= \left(
	 \begin{array}{c}
	 p \\ 2j+3
	 \end{array}
	 \right)~~.\eqno(4.8)$$
We refer to Appendix B for further details, as regards eq.~(4.8).

We are now prepared to discuss the component structure of the NAC deformed
effective action (2.18) in supersymmetric gauge theories. As regards the 
K\"ahler term in eq.~(2.8), we do not have any control of it, so we are going 
to proceed with a generic (undeformed) effective K\"ahler metric 
$$G(\f,\bar{\f})=\pa_{\f}\bar{\pa}_{\bar{\f}}K(\f,\bar{\f})~~.\eqno(4.9)$$
For example, as regards the VY effective action of the N=1 SYM, one has 
\cite{vy}
$$ K(\f,\bar{\f}) =\left(S\bar{S}\right)^{1/3} {}~~~.\eqno(4.10)$$
Only the chiral superpotential $V(f)$ is NAC-deformed according to
eq.~(4.1),  whereas the anti-chiral effective superpotential 
$\bar{V}(\bar{f})$ is not deformed at all. In the VY case, the latter takes 
the same form (2.19).

The bosonic terms contributing to a generic NAC-deformed scalar potential $W$ 
(in components) are thus given
by (we assume that $\VEV{\c^2}=0$) 
$$ -W= G(\f,\bar{\f})M\Bar{M} + \bar{V}'(\bar{\f})\Bar{M}+ 
\fracmm{1}{2 c }\left\{ V\br{\f + c M}-V\br{\f - c M}\right\}~~.
\eqno(4.11)$$
The non-perturbative (algebraic) equations of motion for the auxiliary fields 
$M$ and $\Bar{M}$ are easily solved as
\begin{eqnarray*}
	M &=& -\fracmm{1}{G} \bar{V}'(\bar{\f}),
	\\
	\Bar{M}
	&=&
	-\fracmm{1}{2G}
	\BR{
		V' \br{ \f + c M } 
		+
		V' \br{ \f - c M }
	}        \hfill ~~~~~~~~~~~~~~~~~~~~~~~~~~~~~~~~~~~~~~~~~~~~~(4.12)
	\\
	&=&
	-\fracmm{1}{2G}
	\BR{
		V' \br{ \f - \fracmm{c}{G}  \bar{V}' (\bar{\f}) } 
		+
		V' \br{ \f + \fracmm{c}{G}  \bar{V}' (\bar{\f}) }
	}~.
\end{eqnarray*}

Substituting the solution (4.12) back into eq.~(4.11) gives us the scalar 
potential
$$	W(\f,\bar{\f})
	= 
	\fracmm{1}{2c} 
	\BR{ 
		V \br{ \f + \fracmm{c}{G}  \bar{V}' (\bar{\f}) } 
		-
		V \br{ \f - \fracmm{c}{G}  \bar{V}' (\bar{\f}) }
	}~~.\eqno(4.13)$$
For instance, taking the limit $c\to 0$ in eq.~(4.13) yields the 
standard equation in undeformed supersymmetry,
$$ W_0 = \fracmm{1}{G}V'(\f)\bar{V}'(\bar{\f})~~.\eqno(4.14)$$

The vacuum conditions
$$
\fracmm{ \pa W }{\pa \f } = \fracmm{\pa W }{\pa \bar{\f} } = 0 \eqno(4.15)$$
in the deformed case (4.13) are given by
\begin{eqnarray*}
	\begin{aligned}
	&	\fracmm{1}{G^2} \fracmm{\pa G}{\pa\f} \,\, \bar{V}'(\bar{\f}) 
		\BR{ 
			V' \br{ \f - \fracmm{c}{G}  \bar{V}' (\bar{\f}) } 
			+
			V' \br{ \f + \fracmm{c}{G}  \bar{V}' (\bar{\f}) }
		} 
	\\
	& \qquad \qquad
		+\fracmm{1}{c} 
		\BR{ 
			V' \br{ \f - \fracmm{c}{G}  \bar{V}' (\bar{\f}) } 
			-
			V' \br{ \f + \fracmm{c}{G}  \bar{V}' (\bar{\f}) }
		}
		=0 \hfill  {}~~~~~~~~~~~~~~~~~~~~~~~~(4.16a)
\label{a}
	\end{aligned}
\end{eqnarray*}
and 
$$
		\Br{ 
			\frac{1}{G} \frap{G}{\bar{\f}} \,\, \bar{V}'(\bar{\f}) 
			- \bar{V}''(\bar{\f})
		}\cdot
		\Br{ 
			V'\br{ \f - \fracm{c}{G}  \bar{V}' (\bar{\f})}
	 		+
	 		V'\br{ \f + \fracm{c}{G}  \bar{V}' (\bar{\f}) }
		} =0~~,  \hfill ~~~~~~~~~~~~~(4.16b)
$$
respectively. According to eq.~(4.16b), there are the two possibilities:
\begin{eqnarray*}
	\begin{aligned}
&	\mbox{{\rm case~A}:} \quad
			V'\br{ \f - \fracm{c}{G} \bar{V}' (\bar{\f})}
	 		+
	 		V'\br{ \f + \fracm{c}{G}  \bar{V}' (\bar{\f}) }
			=0~,
&	\mbox{{\rm case~B}:} \quad 		
\fracmm{1}{G} \fracmm{\pa G}{\pa\bar{\f}} \,\, \bar{V}'(\bar{\f}) 
			= \bar{V}''(\bar{\f})~.
	\end{aligned}
\end{eqnarray*}
We now consider those two cases separately.

\noindent \underline{\it The case~A}

Taking into account the remaining equation (4.16a) immediately implies
$$ V' \br{ \f - \fracm{c}{G}  \bar{V}' (\bar{\f}) }=
V' \br{ \f + \fracm{c}{G}  \bar{V}' (\bar{\f}) }=0~~.\eqno(4.17)$$
	
For instance, in the case of the VY superpotential (2.19) with $G=1$ for 
simplicity, one always gets a non-vanishing gluino condensate (2.21) in 
vacuum. As regards the expecation values of the auxiliary fields, we find 
on-shell
$$ M=-\fracmm{1}{G}\bar{V}'(\bar{\f})=\Bar{M}=0~,\eqno(4.18)$$
so that dynamical supersymmertry breaking does not occur. The vacuum 
expectation value of the scalar potential $W$ also vanishes.

\noindent \underline{\it The case~B}

Integration of the differential equation (case B) with respect to the unknown
function $G$ has a general solution 
$$ G (\f,\bar{\f}) =\bar{V}'(\bar{\f}) g(\f)~~,\eqno(4.18)$$
where $g(\f)$ is an arbitrary function. This implies that the on-shell 
K\"ahler metric is a factorizable function of $\f$ and $\bar{\f}$. By the way,
it is the case for the VY K\"ahler potential (4.10). Equations (4.12)
and (4.18) also imply on-shell
$$ M =-\fracmm{\bar{V}'(\bar{\f})}{G}=
 -\fracmm{1}{g(\f)}~~.\eqno(4.19)$$
This means that Dynamical Supersymmetry Breaking (DSB) may be only possible 
when the equation $\bar{V}'(\bar{\f})=0$ has no solutions, just like that in 
the undeformed supersymmetry. Unfortunately, as regards the VY superpotential,
 it is not the case.

\section{Conclusion}

We found that the nilpotent NAC deformation (3.1) gives rise to the simple 
non-perturbative formula (4.1), valid for any chiral superpotential. It is 
worth mentioning that the NAC deformed superpotential (4.1) does not break the 
`Lorentz' invariance since the effective deformation parameter is given by
the scalar (4.2), unlike that at the fundamenal level where the `Lorentz' 
invariance is manifestly broken by $C^{a\b}\neq 0$.

As regards the supersymmetric gauge theories whose NAC deformation is known to
describe some supergravitational contributions, our equation (4.1) is quite 
useful for an explicit computation of the non-perturbative NAC-deformed 
effective actions, though it cannot be used as a tool for further dynamical 
supersymmetry breaking. We provided some physical applications of our equation
 (4.1) to the NAC-deformed supersymmetric Yang-Mills theory described in the 
IR limit by the standard VY superpotential. We found the existence of a
non-vanishing gluino condensate even after the NAC deformation. Unfortunately,
 we also found that no dynamical supersymmetry breaking of the remaining  
$N=1/2$ supersymmetry occurs in the deformed theory.

Possible physical applications of our results are, of course, not limited to
the pure SYM theory. As the most obvious extension, some number of
flavours can be included \cite{2pap}. Since the chiral and anti-chiral 
superpotentials change very differently under the NAC deformation, the 
apparent violation of Hermiticity of the effective action (and, perhaps, of
 its unitarity as well) deserves further study.

\section*{Acknowledgements}

We are grateful to I. Buchbinder, S. Ferrara, A. Ghodsi, R. Ricci and, 
especially, G. Silva and S. Terashima for correspondence.

\newpage

\section*{Appendix A: About our notation}

In this paper we use either Euclidean space with the signature $(+,+,+,+)$,
or Atiyah-Ward space with the signature $(+,+,-,-)$, though we follow the 
standard notation of Wess and Bagger \cite{wb}, invented for supersymmetry in 
Minkowski spacetime. The important differencies are 
emphasized in the main text. Here we merely describe our book-keeping notation
 and the normalization conventions.

The spinorial indices are raised and lowered with the charge conjugation 
matrix that is diagonal in the two-component notation \cite{wb}. Our 
conventions for the two-dimensional Levi-Civita symbol are
$$ \ve_{21}=\ve^{12}=1\quad {\rm so~~that} \quad \ve^{\a\b}\ve_{\b\a}=2~.
\eqno(A.1)$$  

We use the notation
$$ \q\c=\q^{\a}\c_{\a}~,\quad  \q^2= \q^{\a}\q_{\a}=
\ve_{\a\b}\q^{\a}\q^{\b}\eqno(A.2)$$
so that for any two chiral spinors $\q^{\a}$ and $\c^{\a}$ we have
$$\q^{\a}\q^{\b}=-\fracm{1}{2}\ve^{\a\b}\q^2~,\quad 
(\q\c)^2=-\fracm{1}{2}\c^2\q^2~.\eqno(A.3)$$

Our normalization of the Berezin integral over Grassmann coordinates is given 
by
$$ \int d^2\q\,\q^2=1~~.\eqno(A.4)$$
As is well known, Grassmann integration amounts to Grassmann differentiation. 
We use the notation 
$$ \fracmm{\pa^2}{\pa\q^2}= \fracm{1}{4}\ve^{\a\b}\fracmm{\pa}{\pa\q^{\a}}
\fracmm{\pa}{\pa\q^{\b}}~~.\eqno(A.5)$$

The field components $(\f(y),\c(y),M(y))$ of a chiral superfield $f(y,\q)$
are defined by
$$ f= \f +\sqrt{2}\q\c+\q^2 M~,\quad {\rm so~~that}\quad
 \fracmm{\pa^2 f}{\pa\q^2}=M~~.\eqno(A.6)$$ 

The NAC deformation (3.1) is equivalent to the star product (3.5). For 
instance, it is not difficult to check that eq.~(3.5) implies 
$\q^{\a}\star\q^{\b}+\q^{\b}\star\q^{\a}=C^{\a\b}$ indeed. The standard
Grassmann rules for the fermionic coordinates of superspace also get modified 
\cite{sei},
$$ \eqalign{
\q^{\a}\star \q^{\b} ~=~ & -\fracm{1}{2}\ve^{\a\b}\q^2 +\fracm{1}{2}C^{\a\b}~,
\cr
\q^{\a}\star \q^2 ~=~ & C^{\a\b}\q_{\b}~,\cr
\q^2\star \q^{\a} ~=~ & -C^{\a\b}\q_{\b}~,\cr
\q^2\star \q^2 ~=~ & -\det C~.\cr}\eqno(A.7)$$
 
\section*{Appendix B: more about equation (4.8)}

A formal algebraic proof of the identity (4.8) is not very illuminating. So
we offer here another, rather intuitive and easy proof.

\newcommand{\empbox}{
\begin{picture}(10,12)
\put(5,6){\makebox(0,0){\framebox(8,8){}}}
\end{picture}}

\newcommand{\barbox}{
\begin{picture}(10,12)
\put(5,6){\makebox(0,0){\framebox(8,8){}}}
\put(5,6){\makebox(0,0){\rule{3pt}{12pt}}}
\end{picture}}

\newcommand{\pebblebox}{
\begin{picture}(10,12)
\put(5,6){\makebox(0,0){\framebox(8,8){}}}
\put(5,6){\makebox(0,0){$\bullet$}}
\end{picture}}

First, it is not difficult to check that the left-hand-side of eq.~(4.8)
can be rewritten as follows:
$$ \sum_{ \{ \sum k_i = p-j-1\}} 
k_1 k_2 \cdots k_{j+2} \qquad \{ k_i \in \mathbf{N} \} .\eqno(B.1)
$$
The summation is performed over all possible $k_i$'s such that 
$\sum_{i=1}^{j+2} k_i = p-j-1$. For example, in the case of $(j,p)=(1,8)$,
we have
\begin{eqnarray*}
\sum_{ \{ \sum k_i = 6\}} 
k_1 k_2 k_3 &=& 
 1 \cdot 1 \cdot 4 + 1 \cdot 2 \cdot 3 + 1 \cdot 3 \cdot 2 + 
1 \cdot 4 \cdot 1 + 2 \cdot 1 \cdot 3 \nonumber \\
&+& 2 \cdot 2 \cdot 2 + 
2 \cdot 3 \cdot 1 + 3 \cdot 1 \cdot 2 + 3 \cdot 2 \cdot 1 + 
4 \cdot 1 \cdot 1 . \hfill ~~~~~~~~~~~~~~~~~(B.2)  
\end{eqnarray*}

Let's now consider an apparently unrelated problem:\\
{\it
let  $x \! +\! 1$ dots and $x$ sticks line on $n \! + \! x$ small squares
drawn in a row, where $n \! \geq \! x+1$, one item for one square,
provided that each stick is between the dots. In other words, the sticks 
and dots have to appear alternatively on a line. The question is: 
how many ways of their distribution exist?}

We can divide this problem into two steps. Firstly, let's put sticks in such
a way that both neighboring squares of every stick are empty. Secondly, let's
put dots on each breach among sticks one by one.

After putting the sticks that way, there are $n$ empty squares remaining. 
Let $k_i$ be  the numbers of squares in $i$-th breach,
$$
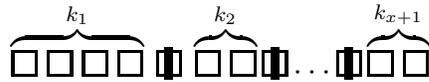
\overbrace{ \empbox \; \empbox \; \empbox \; \empbox}^{k_1} \> \barbox \>
\overbrace{ \empbox \; \empbox }^{k_2} \barbox
\cdots
\barbox \overbrace{ \empbox \; \empbox }^{k_{x+1}}
$$
Due to the condition $k_i \ge 1$ and  the fact that the sum of $k_i$ equals 
to the number of all empty squares,  we have $\sum_{i=1}^{x+1} k_i=n$. 
Once the pattern of sticks is fixed,  the number of ways of putting the 
dots is clearly given by $\prod_{i=1}^{x+1} k_i$. Here is an example
in the case of $(x,n)=(2,6)$:
$$
\begin{array}{ccc}
\pebblebox \empbox \barbox \pebblebox \empbox \empbox \barbox \pebblebox &
\pebblebox \empbox \barbox \empbox \pebblebox \empbox \barbox \pebblebox &
\pebblebox \empbox \barbox \empbox \empbox \pebblebox \barbox \pebblebox \\
\empbox \pebblebox \barbox \pebblebox \empbox \empbox \barbox \pebblebox &
\empbox \pebblebox \barbox \empbox \pebblebox \empbox \barbox \pebblebox &
\empbox \pebblebox \barbox \empbox \empbox \pebblebox \barbox \pebblebox
\end{array}
= 2 \cdot 3 \cdot 1
$$
Actually, one can easily see that this graph exactly corresponds to a term 
in Eq.~(B.2), namely, to $2 \cdot 3 \cdot 1$, while there is 
a one-to-one correspondence between the pattern of sticks and every term there.
  
On the one hand side, when choosing $x+(x+1)$ squares and putting (from the 
left side) a dot, a stick, a dot, etc. alternately on those squares, 
all possible ways of putting sticks and dots appear without repetition. 
Hence, the total number of all patterns is given by the number $\left(
\begin{array}{c} n+x \\ 2x+1 \end{array} \right)$. On the other hand side, 
it is given by a sum of $\prod k_i$, as was mentioned above. Therefore, we 
 get
$$ \sum_{ \{ \sum k_i = n \}} 
k_1 k_2 \cdots k_{x+1} = \left(
\begin{array}{c} n+x \\ 2x+1 \end{array} \right)~. \eqno(B3)$$
It is exactly the relation (4.8) we wanted, after replacing $n=p-j-1$ and 
$x=j+1$.
\vglue.2in

\end{document}